\documentclass[12pt]{article}

\usepackage{graphicx} 
\usepackage{amsmath, amssymb} 
\usepackage{cite} 
\usepackage{caption}       
\usepackage{setspace}
\usepackage{color,soul} 
\usepackage{hyperref} 
\usepackage{xcolor}
\usepackage[numbers,sort&compress]{natbib}
\usepackage[left]{lineno}
\usepackage[bottom=1.2in]{geometry} 

\newcommand{\de}{\mathrm{d}}

\newcommand{\ii}{\mathrm{i}}

\newcommand{\wP}{\omega_\mathrm{p}}
\newcommand{\kB}{k_\mathrm{B}}
\newcommand{\ve}{\varepsilon}

\begin{document}
\begin{center}
  {\LARGE Pathway to Optical-Cycle Dynamic Photonics: Extreme Electron Temperatures in Transparent Conducting Oxides}
  \vspace{1em}

  Jae Ik Choi$^{1,2}$, Vahagn Mkhitaryan$^{2}$, Colton Fruhling$^{2}$, Jacob B. Khurgin$^{3}$,\\
  Alexander V. Kildishev$^{2}$, Vladimir M. Shalaev$^{2}$, Alexandra Boltasseva$^{1,2 *}$
  \vspace{1em}

  {\small
    $^1$School of Materials Engineering, Purdue University, West Lafayette, IN 47907, USA\\
    $^2$Elmore Family School of Electrical and Computer Engineering, Birck Nanotechnology Center, Purdue University, West Lafayette, IN 47907, USA\\
    $^3$Department of Electrical and Computer Engineering, Johns Hopkins University, Baltimore, MD 21218, USA\\[0.5em]
    * aeb@purdue.edu\\
  }
\end{center}

\onehalfspacing

\begin{abstract}
We theoretically demonstrate that transparent conducting oxides (TCOs) exhibit oscillatory and sign-reversing dynamic modulation in transmittance on the order of a few optical cycles under extreme electron temperatures, providing a possible explanation for TCO dynamics observed in earlier experiments. We present an inverse-designed multilayer cavity incorporating an ultrathin TCO layer, which supports an oscillatory optical response more pronounced than those previously observed experimentally in TCOs. This approach yields transmittance oscillations with a characteristic period of $\sim$20\,fs, which corresponds to approximately four optical cycles of the 1.431\,$\mathrm{\mu m}$ probe beam. To achieve a similar oscillatory modulation in $\Delta n$, we incorporate a TCO electron-acceptor layer on top of the inverse-designed cavity, enabling thermionic carrier injection at the TCO junction. The resulting acceptor layer exhibits a striking $\Delta n$ response as fast as 20\,fs, corresponding to only three optical cycles of the 1.8--2.0\,$\mathrm{\mu m}$ probe, and can be further tailored into the sub-optical-cycle regime. The findings could both clarify the previously unexplained transient dynamics in TCOs and, for the first time, demonstrate the critical role of electron temperatures in driving oscillatory dynamic responses. 
\end{abstract}

\section{Introduction}\label{sec1}
The ability to control light–matter interactions in the time domain has far-reaching implications for various emerging applications such as quantum computing~\cite{Alu_PTC}, super-resolution imaging~\cite{theory_application_PTC}, ultralow-noise optical devices~\cite{Alu_PTC}, thresholdless lasers~\cite{theory_application_PTC}, and the more fundamental capability of quantum field control \cite{free_electron_PTC, theory_application_PTC}. Photonic time crystals, in which the refractive index is periodically modulated in time rather than in space, represent a promising paradigm for such applications. However, despite these compelling motivations, experimental realization of photonic time crystals in the visible to infrared spectrum has not yet been observed. The main challenge lies in the requirement for extremely large refractive index modulation on the timescale of an optical cycle, which no photonic materials have achieved to date~\cite{Vlad_Segev_PTC, theory_application_PTC, Narimanov_TCO}.

To this end, transparent conducting oxides (TCOs) have emerged as one of the most promising material platforms for time-varying photonics, including photonic time crystals. Prior studies have demonstrated that TCOs can exhibit a near-unity change in refractive index with a full relaxation time of less than 500\,fs~\cite{Boyd_TCO_science, Kinsey_AZO_optica}. Such a response is both several orders of magnitude faster and stronger than noble metals, which exhibit only picosecond-scale modulation at best. The distinct modulation strength in TCOs arises from their unique nonlinearities associated with the non-parabolic conduction band, and the epsilon-near-zero (ENZ) enhancement, where the real part of permittivity vanishes and produces a strong field enhancement and a slow light effect~\cite{Boyd_TCO_science, Kinsey_AZO_optica, Boyd_TCO_Nature_photonics, Jaffray_review, Boyd_higer_nonlinear, Kinsey_Khurgin_nonlinear, kinsey_NZI_review, Boyd_nonlinear_review}. The superior response time in TCOs, meanwhile, generally originates from the combination of strong electron-phonon coupling and low electron heat capacity, which enables rapid heating and cooling of the electron temperature~\cite{LEDD, Khurgin_fast_slow, Kinsey_ENZ_nonlinear_review, Nanophotonics_scattering}.

However, in our recent study, we reported an even more striking phenomenon: a transient transmission decay and a spectral shift in the probe signal occurring on a near-optical cycle timescale~\cite{Moti_TCO_experiment}. While this behavior lies well outside the conventional electron-phonon relaxation regime and remains physically unexplained, it was consistently observed across a range of pump pulse durations, suggesting the existence of a robust and reproducible mechanism. Such rapid responses in the early stage of a dynamic process were observed in a few studies, while largely unexamined and received limited attention despite their potential significance~\cite{Chang_Nature_photonics, initial_dynamics_1, initial_dynamics_2}. Taken together, these observations further underscore the potential of TCOs as distinctly compelling materials for emerging time-varying photonic platforms. 

These unprecedented femtosecond-scale phenomena are closely connected to the unique abilities of TCOs to reach electron temperatures that approach---or even exceed---the Fermi temperature under optical excitations. In this unique domain of low Fermi energy and carrier density $n_\mathrm{e}$ combined with extreme electron temperatures, ($n_\mathrm{e}\lambda_T^3 \ll 1$, where $\lambda_T=\sqrt{2\pi\hbar^2/m^* \kB T}$ is the thermal de Broglie wavelength)~\cite{LEDD, Khurgin_fast_slow, Kinsey_ENZ_nonlinear_review}, the Fermi-Dirac distribution begins to resemble the classical Maxwell-Boltzmann profile, leading to a partial breakdown of quantum degeneracy~\cite{FD_MB}. As electrons enter this fundamentally different regime, they behave as a classical plasma-like ensemble, with enhanced interaction with the phonons, impurities, and grain boundaries that can lead to unconventional carrier dynamics. Inspired by this distinct behavior, we explore the transient nonlinear optical behavior of TCOs under such extreme electron temperature conditions, revealing the pivotal role of hot-electron interaction in shaping their femtosecond-scale non-trivial dynamics.

For this purpose, we utilize the two-temperature model that accounts for the thermalized-electron dynamics. Importantly, our simulations could explain the previously observed femtosecond-scale relaxation of the transmission dynamics reported for indium tin oxide (ITO) in Ref.~\cite{Moti_TCO_experiment}. This provides a reasonable physical explanation for the observed non-trivial dynamics and validates the predictive capability of our framework (Supplementary Section~1). To explore pathways for engineering nontrivial dynamic responses on shorter timescales, we use this simulation framework to study the TCO response under extreme electron-temperature conditions enabled by the inverse-designed cavity. Next, we investigate whether a similar oscillatory response in the refractive index can be achieved using a Sommerfeld model representing a hypothetical carrier-injection scenario. Finally, guided by the Sommerfeld-model-based hypothetical scenario, we propose an experimentally feasible multilayer cavity incorporating a TCO bilayer with engineered carrier densities and thicknesses, where interfacial carrier injection is facilitated under optical excitation. Together, we aim to establish a promising mechanism and material platform for achieving strong optical modulations on optical-cycle timescales, paving the way for emerging dynamic photonic applications operating in the infrared to visible range.

\section{Design \& Theory}\label{sec2}

\subsection*{Boosting local electron temperature via inverse-designed ENZ cavity}

\begin{figure}[htbp]
    \centering
    \includegraphics[width=1\textwidth]{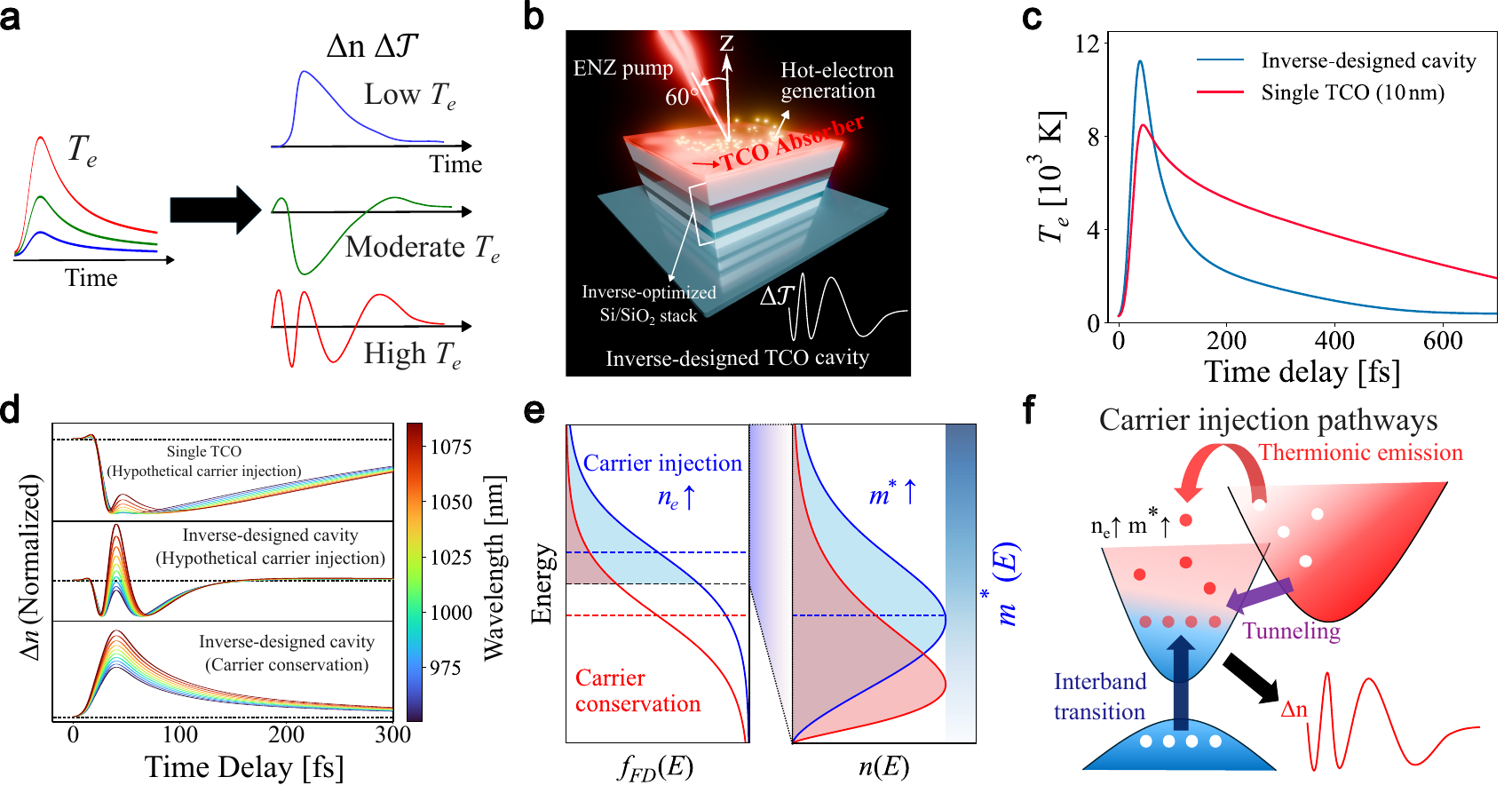}
    \caption{
    \textbf{Oscillatory dynamics in the inverse-designed cavity.}
    \textbf{a}, Illustration of oscillatory dynamics for different peak electron temperatures. Higher peak electron temperature results in a dynamic response evolving into a more complex oscillatory behavior.  
    \textbf{b}, Schematic of the inverse-designed cavity, optimized for maximizing the electron temperature in a 10\,nm TCO layer. The multilayer stack is simulated under excitation by a 10\,fs, $p$-polarized pump pulse centered at the ENZ wavelength of 1425\,nm and incident at 60$^\circ$, with a fluence of 2\,mJ/cm$^2$. 
    \textbf{c}, Simulated electron temperature evolution in the 10\,nm TCO layer, revealing enhanced electron temperature in the inverse-designed cavity compared to a single TCO film on a fused silica substrate. 
    \textbf{d}, Normalized refractive index modulation ($\Delta n$) dynamics in the TCO, comparing different geometry and chemical potential models. The pump excitation conditions are identical to those used in panel b. Top: response of 10\,nm single TCO/fused silica under the hypothetical carrier injection. Middle:  Oscillatory $\Delta n$ response in the cavity-integrated TCO film with the hypothetical carrier injection. Bottom: Single-peaked $\Delta n$ response of the cavity-integrated TCO film under a carrier conservation.
    \textbf{e}, Illustration of the mechanism driving $\Delta n$ oscillations in the high $T_\mathrm{e}$ carrier injection scenario. Simultaneous increase in $n_\mathrm{e}$ and $m^*$ shifts the plasma frequency in opposite directions, giving rise to an oscillatory $\Delta n$ response. Left: Fermi-Dirac distribution for Sommerfeld versus carrier-conserving models. Right: Normalized energy distribution of total carriers, highlighting a large high-energy carrier population and increased effective mass in the hypothetical carrier injection (Sommerfeld) case.
    \textbf{f}, Schematic of representative electron injection pathways in TCOs for achieving dynamic carrier-density modulation, illustrated on a non-parabolic band structure to highlight the competing roles of $n_\mathrm{e}$ and $m^*$.
    }
  \label{fig:fig_1}
\end{figure}

Extreme electron temperatures in TCOs can open a new regime of time-varying photonics, where non-trivial dynamics offer dramatically faster modulation features. Figure~\ref{fig:fig_1}(a) illustrates the distinct modulation behavior addressed in this work: an oscillatory behavior of the refractive index and transmittance modulation that emerges at high electron temperatures. While a conventional single-peaked response is observed at moderate excitation powers in TCOs, increasingly complex oscillatory dynamics can emerge under TW/cm$^2$-level excitation. Importantly, such an oscillatory response can play a central role in reaching optical-cycle modulation speeds. The oscillation subdivides the overall relaxation process, spanning several hundred femtoseconds, into multiple intervals, thereby yielding substantially shorter temporal periods within each interval. We aim to establish a theoretical understanding of how these non-trivial responses emerge at extreme electron temperatures by modeling and comparing the dynamic behavior of TCOs in two distinct configurations: an inverse-designed optical cavity and a single film on a fused silica substrate. 

To boost the local electron temperatures within the TCO layer and induce oscillatory dynamic responses, we inverse design a multilayer optical cavity that enables strong light absorption in an ultrathin TCO film. Given the target absorptance spectrum, the inverse-design algorithm, which incorporates the transfer matrix-based optical multilayer solver with gradient backpropagation, optimizes the Si/SiO$_2$ layer thicknesses and carrier concentration of the ultrathin TCO layer~\cite{Caldwell_inverse}. The high absorptance at the targeted wavelength ($\rm \lambda=1425\,nm$) is achieved through strong field confinement enabled by the cavity-enhanced ENZ resonance of the multilayer cavity. The structure consists of a 10\,nm TCO film on top of a Si/SiO$_2$-based dielectric mirror with five layers of varying thicknesses (Figure~\ref{fig:fig_1}(b)). The complex permittivity of the TCO film is described using a Drude model parameterized with background permittivity, carrier density-dependent plasma frequency, and Drude damping obtained from an aluminum-doped zinc oxide (AZO) film. The optimized carrier concentration is $7.35 \times 10^{20}\,\mathrm{cm}^{-3}$ with the corresponding complex permittivity presented in Supplementary Figure~3(a). Although AZO is used here as a reference material, the design framework is equally compatible with other TCOs and tunable Drude-type materials such as indium tin oxide (ITO), gallium-doped zinc oxide (GZO), cadmium oxide (CdO), transition-metal nitrides (TMNs), and MXenes. Remarkably, the designed cavity structure achieves over $60\%$ absorptance at the ENZ wavelength for \(p\)-polarized incident light, even within the ultrathin thickness of only 10\,nm. This is a threefold enhancement compared to the single 10\,nm TCO film on fused silica with identical optical properties (Supplementary Figure~3(c)). Hereafter, we refer to the 10\,nm TCO layer in the cavity as the absorber layer, as it primarily accounts for the strong optical absorption.

Upon ultrafast optical excitation of the inverse-designed cavity at the ENZ wavelength, strong optical confinement within the TCO layer drives a pronounced intraband electronic response. While interband transitions may occur through multiphoton absorption, the relevant interband absorption of TCO materials typically lies in the ultraviolet, around 300--350\,nm, requiring approximately four to five-photon absorption under the close-to-ENZ excitation wavelength considered here~\cite{interband_AZO_IOP}. Because the probability of this process is extremely low, we assume that the collective intraband response dominates the nonlinear dynamics in the TCO layer. We calculate the resulting electron-temperature dynamics using a spatially resolved two-temperature model (Supplementary Section~5). The pump pulse is described by a Gaussian temporal envelope with a 10\,fs pulse duration and ENZ central wavelength. After Fourier transforming the temporal profile into the frequency domain, we calculate the absorbed energy for each spectral component using the transfer-matrix method, thereby providing spatially resolved absorption within the multilayer stack. This absorbed-energy profile serves as the input heat source for the coupled two-temperature equations, from which we obtain the spatio-temporal evolution of the electron temperature. This spatially resolved framework is particularly advantageous for multilayer systems because it preserves the spatial dependence of optical absorption and the corresponding thermal electron dynamics, which is often neglected in simplified treatments. Such spatial information is especially important at TCO-semiconductor interfaces, where electron transport across the junction critically influences the overall transient response. In particular, hot electrons can undergo substantial thermionic emission across the junctions at elevated electron temperatures~\cite{Thermionic_Sim, Thermionic_Sim2, Thermionic_Sim３}. In addition, key material parameters that determine the dynamics, including the chemical potential, electron heat capacity, and electron-phonon coupling, depend nonlinearly on the electron temperature~\cite{Boyd_TCO_science, Chang_Nature_photonics, Boyd_TCO_Nature_photonics, TTM_FDTD}. To capture these effects, we incorporated an interface condition that accounts for thermionic emission across the TCO/Si junction, along with electron temperature-dependent thermophysical properties into the two-temperature model (details in Supplementary Section~5, 6). The Si/SiO$_2$ interface in the cavity structure was considered insulating due to the high energy barrier and low electron temperature in the Si layer.

Figure~\ref{fig:fig_1}(c) shows the simulated evolution of the electron temperature in the TCO layer for both the single film on fused silica and the inverse-designed cavity. In the numerical simulation, the structure is excited by a 10\,fs, \(p\)-polarized pump at ENZ central wavelength (1425\,nm), with 2\,mJ/cm$^2$ fluence and 60$^\circ$ grazing angle (see Figure~\ref{fig:fig_1}(b)). This corresponds to a peak intensity of approximately 200\,GW/cm$^2$, remaining well below the TW/cm$^2$-level intensities typically employed in ultrafast TCO studies and therefore remains within a realistic excitation regime~\cite{Moti_TCO_experiment, Kinsey_ENZ_nonlinear_review}. The ENZ pump wavelength and 60$^\circ$ incidence angle are chosen to maximize the electron temperature change in TCOs, owing to the strongly enhanced absorption of $z$-field component in the ENZ regime. We note that achieving the same electron temperature with a non-ENZ pump centered at 800\,nm requires a fluence more than two orders of magnitude higher, underscoring the effectiveness of ENZ excitation in enhancing the thermal electron response. As a result, the cavity-integrated TCO film exhibits enhanced electron temperature compared to a single TCO layer on fused silica, reaching values close to the Fermi temperature ($T_\mathrm{F}=11584\,\mathrm{K}$). Although absorption is enhanced by nearly threefold in the cavity-integrated TCO layer (Supplementary Figure~3(c)), the resulting increase in electron temperature is moderated by the rise in electronic heat capacity at elevated electron temperatures. Nevertheless, as the electron temperature approaches the Fermi temperature, thermal broadening of the electron distribution in the conduction band becomes sufficient to substantially alter hot-carrier dynamics, leading to markedly different transient modulation behavior (Figure~\ref{fig:fig_1}(d)).

\subsection*{Insights from a hypothetical carrier-injection model for optical-cycle-timescale $\Delta n$}
We begin this section by introducing a temperature-dependent model for the refractive index. A key component of this model is the treatment of the chemical potential, which can be evaluated either using the Sommerfeld approximation or a more rigorous carrier-conserving approach. We examine both cases to clarify how the realistic carrier-conserving response differs from the approximate Sommerfeld case, in which the effective carrier contribution varies with electron temperature. Although the Sommerfeld approximation does not strictly represent the realistic carrier-conserving response, it provides important insights into the conditions required to achieve sign-reversing oscillations in refractive index $\Delta n$ modulation. Therefore, the Sommerfeld model provides design guidance for engineering a more realistic structure that supports oscillatory $\Delta n$ dynamics. Guided by this insight, we later propose a more realistic structure that extends the carrier-conserving model by coupling with thermionic carrier-injection physics to enable oscillatory $\Delta n$ dynamics.

To capture the dynamic response across different electron temperatures in a TCO film, we calculate the time-dependent refractive index change $\Delta n$ from the electron temperature-dependent Drude formula (Eq.~\ref{eq:updated_Drude_index}), where both the plasma frequency ($\wP$) and Drude damping ($\Gamma$) arising from electron scattering with phonon, impurity, grain-boundary, and surface roughness are dynamic functions of the electron temperature~\cite{Boyd_TCO_science, Chang_Nature_photonics, Boyd_TCO_Nature_photonics}. These dependencies are given by,
\begin{equation}
n(\omega, T_\mathrm{e}) = \sqrt{\ve(\omega, T_\mathrm{e})} = \sqrt{\ve_{\infty} - \frac{\wP(T_\mathrm{e})^2}{\omega^2 + \ii \omega \Gamma(T_\mathrm{e})}},
\label{eq:updated_Drude_index}
\end{equation}
\begin{equation}
\wP^2(T_\mathrm{e}) = \frac{e^2}{3 m^\ast \ve_0 \pi^2} \int_{0}^{\infty} \frac{\de E}{1 + 2 C E} \left( \frac{2m^*}{\hbar^2 (E + C E^2)} \right)^{3/2} \left( -\frac{\partial f_\mathrm{FD}(E, \mu(T_\mathrm{e}), T_\mathrm{e})}{\partial E} \right),
\label{eq:plasma_update}
\end{equation}
\begin{equation}
\Gamma(T_\mathrm{e}) = \Gamma_0 \left[ 1 + \left(\frac{T_\mathrm{e}}{T_0}\right)^m \right],
\label{eq:damping_update}
\end{equation}
where $f_{\mathrm{FD}}$ denotes the Fermi--Dirac distribution, $\mu(T_\mathrm{e})$ is the electron temperature-dependent chemical potential, $C$ characterizes the degree of band non-parabolicity, $\Gamma_0$ is the Drude damping at room temperature, $m$ describes the power-law dependence of the damping on $T_\mathrm{e}$, and $T_0$ is a fitting parameter that sets the characteristic temperature scale of this dependence. Because both $m$ and $T_0$ depend sensitively on the sample-specific microstructure, their values must be calibrated against experimental measurements.

Earlier models underestimated the influence of electron temperature on Drude damping because the small Fermi surface in TCOs suppresses Umklapp processes, thereby limiting the direct contribution of electron-electron scattering to optical absorption. More recent studies, however, reveal that the damping rate can increase far more strongly with electron temperature than previously assumed. In fact, the damping-induced contribution to the optical response can become comparable in magnitude to that arising from the plasma frequency shift, making enhanced damping a central driver of the nonlinear optical response at high electron temperatures~\cite{Sivan_loss, LEDD, LEDD_rigorous}. This strong dependence originates from rapid electron heating, which increases the momentum-relaxation rate via electron-phonon, impurity, grain-boundary, and surface-roughness scattering~\cite{polycrstal_scattering, ab_initio_ep, LEDD}. While previous studies assumed an approximately linear increase at moderate temperatures, the theoretical models in Refs.~\cite{Sivan_loss, LEDD, LEDD_rigorous} and the experimental fitting reported in Refs.~\cite{Nanophotonics_scattering, Scattering_linear} predict a much steeper dependence at higher electron temperatures. Guided by these findings, we adopt a quadratic temperature dependence $m=2$ as the baseline model in this work, while also performing a sensitivity analysis over a range of $m$ values. The parameter $T_0$ is then determined by fitting the model to the measured non-trivial relaxation dynamics reported in Ref.~\cite{Moti_TCO_experiment} (Supplementary Section~1). The resulting formulation could explain the previously observed relaxation behavior, as detailed in the simulation-experiment comparison in Supplementary Section~1.

To complete the electron temperature-dependent Drude description, we next calculate the chemical potential of the thermalized electron distribution using two distinct models. The first is the Sommerfeld model.
\begin{equation}
\mu(T_\mathrm{e}) = \varepsilon_F \left[ 1 - \frac{\pi^2}{12} \left( \frac{T_\mathrm{e}}{T_\mathrm{F}} \right)^2 \right],
\label{eq:Sommerfeld}
\end{equation}
which we use in this work to represent a hypothetical case with dynamic carrier-density modulation. Here, $\varepsilon_F$ is the Fermi energy, and $T_\mathrm{F}$ is the Fermi temperature. Although the Sommerfeld approximation becomes inaccurate at high electron temperatures by predicting an unphysical increase in carrier density~\cite{Thermionic_Sim}, comparison with the carrier-conserving model provides useful design intuition. In particular, it clarifies the role of dynamic carrier-density modulation, the extent of carrier injection required, and the physical mechanism underlying the oscillatory $\Delta n$ dynamics. The Sommerfeld model is therefore used only in Figure~\ref{fig:fig_1}(d) and Figure~\ref{fig:fig_2}(a), where it is labeled as hypothetical carrier injection, whereas all other results rely on the more realistic carrier-conserving model.

The more realistic carrier-conserving model is obtained by solving the integral equation,
\begin{equation}
n_{e} = \int_{0}^{\infty} f_\mathrm{FD}(E,\mu(T_\mathrm{e}),T_\mathrm{e}), g(E),\mathrm{d}E,
\label{eq:conservation}
\end{equation}
which enforces carrier conservation under intraband optical excitation. Here, $n_e$ is the carrier density, $f_\mathrm{FD}$ denotes the Fermi-Dirac distribution, and $g(E)$ is the electronic density of states.

To identify the conditions required for an oscillatory $\Delta n$ response, we compare the refractive index dynamics of TCOs across different geometries and chemical potential models using Eqs.~(\ref{eq:updated_Drude_index}--\ref{eq:conservation}), as shown in Figure~\ref{fig:fig_1}(d). Oscillatory $\Delta n$ appears only in the hypothetical carrier-injection case for the inverse-designed cavity, where cavity enhancement combined with ENZ pumping drives the electron temperature to extreme values. In particular, the single TCO film (top) exhibits no pronounced oscillatory behavior, apart from an initial weak positive peak followed by a subsequent sign reversal. In contrast, the cavity-integrated TCO film (middle) displays pronounced multi-period oscillations. This difference arises because the cavity resonance produces a stronger electron temperature rise. When carrier conservation is imposed, however, the $\Delta n$ response reverts to a non-oscillatory form even in the cavity geometry (bottom). For reference, a single 10\,nm TCO film on fused silica under non-ENZ excitation (800\,nm pump, Supplementary Figure~7) also shows no sign reversal because the electron temperature remains low in the absence of ENZ enhancement.

The physical mechanism underlying these unique oscillations in the Sommerfeld model can be understood by examining the conduction-band electron density and energy distribution at elevated electron temperatures. Figure~\ref{fig:fig_1}(e) presents the corresponding Fermi-Dirac distributions for the Sommerfeld (blue) and carrier-conserving (red) cases. The black dotted line marks the conduction-band edge, whereas the blue and red dotted lines denote the corresponding chemical potentials. The markedly larger area under the blue curve indicates that the Sommerfeld approximation overestimates the total carrier density as the electron temperature rises (Supplementary Figure~12(a)), thereby effectively representing the behavior of the hypothetical carrier-injection scenario. The normalized electron density, $n_\mathrm{e}(E)=f_\mathrm{FD}\cdot g(E)$, shown in the right-hand panel of Figure~\ref{fig:fig_1}(e), further reveals a larger average effective mass in the carrier-injection case owing to the greater electron occupation of high-energy states. The shaded blue bar on the right highlights the increase in effective mass with energy that arises from the non-parabolic band structure. Together, electron-temperature-controlled carrier injection in the hypothetical carrier-injection scenario introduces two competing contributions to the plasma frequency: an increase in carrier density raises the plasma frequency, whereas an increase in average effective mass reduces it. Because these contributions shift the plasma frequency in opposite directions (Eq.~\ref{eq:plasma_update}), they induce multiple $\Delta n$ sign reversals and ultimately give rise to an oscillatory response. Importantly, this behavior becomes pronounced only when the electron temperature is sufficiently high, such that the changes in carrier density $n_\mathrm{e}$ and effective mass $m^*$ become large enough to drive competing contributions to the plasma-frequency modulation.

This form of oscillatory modulation is particularly pronounced in TCOs because their optical response occupies a unique regime between conventional semiconductors and metals. In conventional semiconductors, the plasma contribution to the permittivity is typically weak, whereas in metals, the high carrier density limits the rapid electron heating and large carrier-density-driven optical tunability. TCOs, by contrast, combine a substantial plasma-frequency shift $\wP$ and damping modulation $\Gamma$, both of which strongly alter the permittivity, with strong electron-phonon coupling and ENZ-enhanced light-matter interaction. Together, these features enable an unusually strong, rapid, and qualitatively distinct dynamic optical response.

The distinctive properties of ENZ TCOs that enable ultrafast oscillatory optical modulation can be summarized as follows. First, their intermediate carrier density is sufficiently high to strongly modulate the optical properties via the plasma response, yet low enough to allow rapid electron heating, so the electron temperature efficiently approaches the Fermi temperature. Second, the non-parabolic conduction band causes carriers driven to higher energies to acquire a substantially larger effective mass, thereby producing a pronounced modification of the plasma frequency. Third, the Drude damping rate $\Gamma$ in TCOs exhibits an unusually strong dependence on electron temperature through their distinct electron-scattering mechanisms, in contrast to the weaker damping variation typically observed in conventional semiconductors. This establishes a competition between $\wP$ and $\Gamma$ that can give rise to an oscillatory optical response. Fourth, full electron-energy relaxation through phonons is exceptionally rapid in TCOs because of their high phonon energies, enabling complete relaxation within a sub-picosecond timescale. This is orders of magnitude faster than in conventional metals and interband-transition-driven responses in semiconductors. Finally, these light-induced effects are further enhanced under the ENZ condition, while probing near the ENZ wavelength provides exceptional sensitivity to changes in the optical response, reflected in the large relative permittivity change $\Delta \varepsilon / \varepsilon$.

Building on these considerations, Figure~\ref{fig:fig_1}(f) highlights several distinct routes that can, in principle, enable concurrent modulation of $n_\mathrm{e}$ and $m^*$. The conduction-band illustration depicts a heterointerface, where electrons can be spatially transported across the junction. These routes include electrical gating in heterostructures to induce tunneling~\cite{Gating, AZO_gating, AZO_VO2_gating} and spatial hot-carrier transport processes~\cite{Graphene_hot_carrier, Khurgin_hot_carrier, Khurgin_hot_electron, Khurgin_hot_electron2}. Another possible route is interband excitation~\cite{interband_Ncomm, interband_scientific_reports, interband_AZO_IOP}, which is achievable within a single material but generally requires stronger optical excitation to induce an oscillatory response. Note that the interplay between intraband and interband excitations in ITO could lead to oscillatory transmission (reflection) dynamics~\cite{Mustafa_TCO_dynamics}, which was first observed in~\cite{Moti_CLEO_2025}, but the underlying mechanism remained unexplained. However, to achieve appreciable control over $n_e$ modulation, spatial hot-carrier injection via thermionic emission is particularly attractive because of its strong dependence on electron temperature and intrinsically ultrafast character~\cite{Thermionic_space_charge, WS2_injection, MC_Gold_injection, Nanocube_injection}. In later sections, we examine oscillatory $\Delta n$ in a more realistic and experimentally accessible thermionic carrier-injection platform based on a TCO bilayer integrated into a multilayer optical cavity.

\subsection*{Sign-map analysis of oscillatory $\Delta n$ and $\Delta \mathcal{T}$ responses}

\begin{figure}[htbp]
    \centering
    \includegraphics[width=1\textwidth]{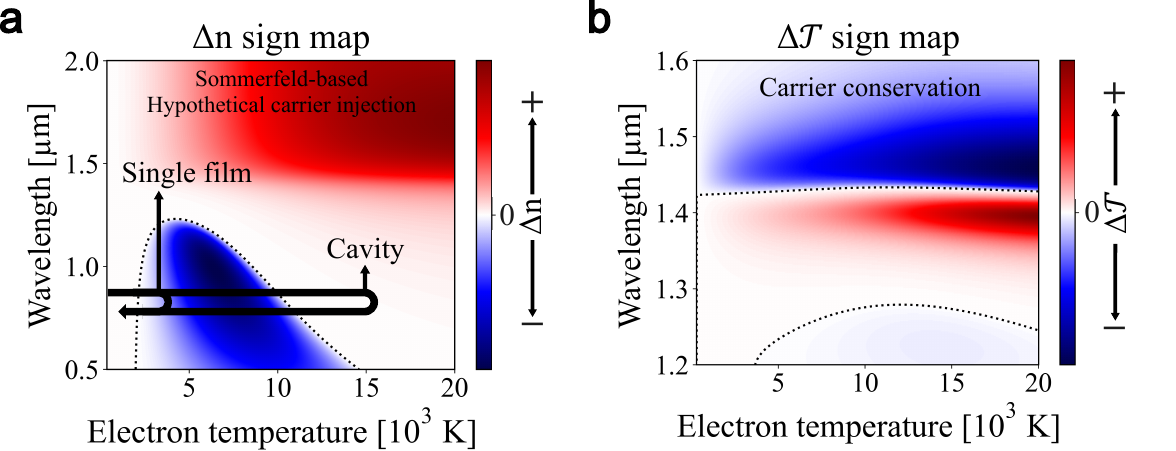}
    \caption{
    \textbf{Electron temperature-dependent $\Delta n$, $\Delta \mathcal{T}$ sign map.}  
    \textbf{a}, Sign map of $\Delta n$ as a function of electron temperature and probe wavelength under hypothetical carrier injection. Red and blue denote positive and negative $\Delta n$ regions, respectively. The arrows highlight the role of electron temperature in enabling the complex oscillatory $\Delta n$ response, which can become strongly oscillatory in cavity-integrated TCO films.
    \textbf{b}, Sign map of $\Delta \mathcal{T}$, revealing two distinct spectral regions of sign reversal. Unlike $\Delta n$, oscillations in $\Delta \mathcal{T}$ do not require carrier-density modulation and are therefore modeled under carrier conservation.
    }
    \label{fig:fig_2}
\end{figure}

In the previous section, we established that oscillatory $\Delta n$ requires both high electron temperatures and dynamic carrier-density modulation. To identify the electron-temperature regime in which multiple sign reversals emerge and become increasingly complex, we construct a sign map by plotting $\Delta n$ as a function of electron temperature and probe wavelength. In this analysis, the chemical potential is evaluated using the Sommerfeld formulation (Eq.~\ref{eq:Sommerfeld}), which implicitly represents a hypothetical carrier injection scenario in TCOs. The resulting $\Delta n$ sign map in Figure~\ref{fig:fig_2}(a) provides a conceptual picture of how the oscillatory temporal response of the TCO film can be finely controlled by engineering the electron temperature response and probe wavelength. A more experimentally accessible realization of this concept is introduced later through the thermionic carrier injection scheme.

The $\Delta n$ maps in Figure~\ref{fig:fig_2}(a) reveal the conditions under which sign reversals occur. The dotted zero-crossing line, which separates the positive and negative $\Delta n$ regions, marks the electron temperatures at which the modulation changes sign. The number of oscillation periods is determined by how many times the electron temperature trajectory during electron heating and relaxation (black arrows) intersects this zero-crossing line. For a single TCO film on fused silica under moderate pump powers, the peak electron temperature can reach the value indicated in the figure, and $\Delta n$ crosses the zero-crossing boundary twice. By contrast, in the cavity configuration, where absorption is strongly enhanced, the electron temperature can reach substantially higher values. As a result, the trajectory intersects the zero-crossing boundary more frequently, crossing both sides of the negative ($\Delta n<0$, blue) region, producing four sign reversals and a correspondingly more complex oscillatory response. Consequently, the temporal response changes substantially depending on the peak electron temperature reached (Figure~\ref{fig:fig_1}(a)), indicating that higher electron temperatures favor more oscillatory dynamics.

Figure~\ref{fig:fig_2}(b) shows a similar sign map in the transmittance modulation $\Delta \mathcal{T}$, calculated by the following formula: 
\begin{equation}
\ln\left(\frac{\mathcal{T}_\mathrm{mod}(\lambda, T_\mathrm{e})}{\mathcal{T}(\lambda)}\right) = \frac{\partial \ln \mathcal{T}(\lambda)}{\partial \varepsilon'} \Delta \varepsilon'(T_\mathrm{e}) + \frac{\partial \ln \mathcal{T}(\lambda)}{\partial \varepsilon''} \Delta \varepsilon''(T_\mathrm{e}),
\label{eq:del_T}
\end{equation}
where $\mathcal{T}_\mathrm{mod} = \mathcal{T} + \Delta \mathcal{T}(T_\mathrm{e})$, and the real and imaginary permittivity change is modeled using the electron temperature-dependent Drude formulation (Eq.~\ref{eq:updated_Drude_index}), while enforcing carrier conservation under intraband excitation (Eq.~\ref{eq:conservation}).

Carrier conservation is imposed in the transmittance-modulation analysis because the oscillatory responses of $\Delta n$ and $\Delta \mathcal{T}$ fundamentally originate from different physical mechanisms. Unlike oscillatory $\Delta n$, oscillatory $\Delta \mathcal{T}$ does not require carrier-density modulation. It is therefore important to distinguish between these two responses, as oscillations in $\Delta \mathcal{T}$ can easily be misconstrued as evidence of oscillatory $\Delta n$. Sign changes in $\Delta \mathcal{T}$ arise from the competition between the real and imaginary parts of the complex permittivity (Eq.~\ref{eq:del_T})---or, similarly, from the competition between $\Delta \omega_\mathrm{p}$ and $\Delta \Gamma$---whereas sign reversals in $\Delta n$ originate from competing contributions of $n_\mathrm{e}$ and $m^*$. In fact, $\Delta \mathcal{T}$ oscillations are associated with sign reversals in the extinction coefficient ($\Delta k$) even when $\Delta n$ remains single-peaked (Supplementary Section~10). Although such $\Delta \mathcal{T}$ sign reversals have been reported previously~\cite{Moti_TCO_experiment, Chang_Nature_photonics, initial_dynamics_1, initial_dynamics_2}, achieving sign reversal in $\Delta n$ is considerably more demanding because it requires substantial carrier density modulation. In other words, oscillations in the real part of the complex refractive index ($\Delta n$) are far more difficult to realize than oscillations in the imaginary part ($\Delta k$).

The $\Delta \mathcal{T}$ sign map (Figure~\ref{fig:fig_2}(b)) exhibits two distinct zero-crossing regions (dotted lines). Similar to the $\Delta n$ map in Figure~\ref{fig:fig_2}(a), the electron temperature trajectory can cross these zero-crossing boundaries multiple times when sufficiently high values are reached. Although the relevant spectral window is relatively narrow, this gives rise to oscillatory $\Delta \mathcal{T}$ in two distinct spectral regions, near and away from the ENZ wavelength. Overall, the sign-map analysis indicates that extreme electron temperatures are essential for producing oscillatory temporal responses in both $\Delta n$ and $\Delta \mathcal{T}$, thereby suggesting extreme electron temperatures as a route toward optical-cycle-timescale modulation of the dynamic optical response in TCOs.

\section{Results}\label{sec3}
\subsection*{Optical-cycle-timescale $\Delta \mathcal{T}$ oscillations}

In this section, we investigate the transmittance dynamics of the inverse-designed cavity and clarify the origin of transmittance oscillations occurring on a few optical cycle timescales. The transient response can be understood through Eq.~\ref{eq:del_T}. Using the time evolution of the electron temperature from the two-temperature model (Supplementary Section~5, 6), we obtain the relative change of transmittance $\Delta \mathcal{T}/\mathcal{T}$ as a function of the pump-probe delay (Eq.~S1). This formulation provides a robust framework for capturing the dynamic response of the system and for reliable interpretation of the observed transient changes~\cite{Ncomm_del_T_Eq, TPP_modeling}.

We define $\alpha \equiv {\partial \ln \mathcal{T} (\lambda)}/{\partial \varepsilon'}$ and $\beta \equiv {\partial \ln \mathcal{T}(\lambda)}/{\partial \varepsilon''}$, and refer to them as the refractive (propagation-related) and absorptive sensitivities, respectively. The refractive sensitivity $\alpha$ reflects changes in light propagation induced by electron-energy redistribution within the conduction band under intraband excitation, whereas the absorptive sensitivity $\beta$ describes the increase in Drude loss associated with enhanced carrier scattering. Together with the real and imaginary permittivity variations, these sensitivities quantify the respective refractive ($\alpha \Delta \varepsilon'$) and absorptive ($\beta \Delta \varepsilon''$) contributions to the transmittance response. Because $\alpha$ and $\beta$ vary spectrally according to the cavity resonance profile (Supplementary Figure~6), a wavelength range emerges in which their signs are opposite. In this regime, the refractive and absorptive contributions drive the optical response in competing directions, toward either a more dielectric (transmissive) or more metallic (absorptive) state. This competition gives rise to oscillatory transmittance modulation with sign reversals, depending on which contribution dominates at a given time and probe wavelength.

\begin{figure}[htbp]
    \centering
    \includegraphics[width=1\textwidth]{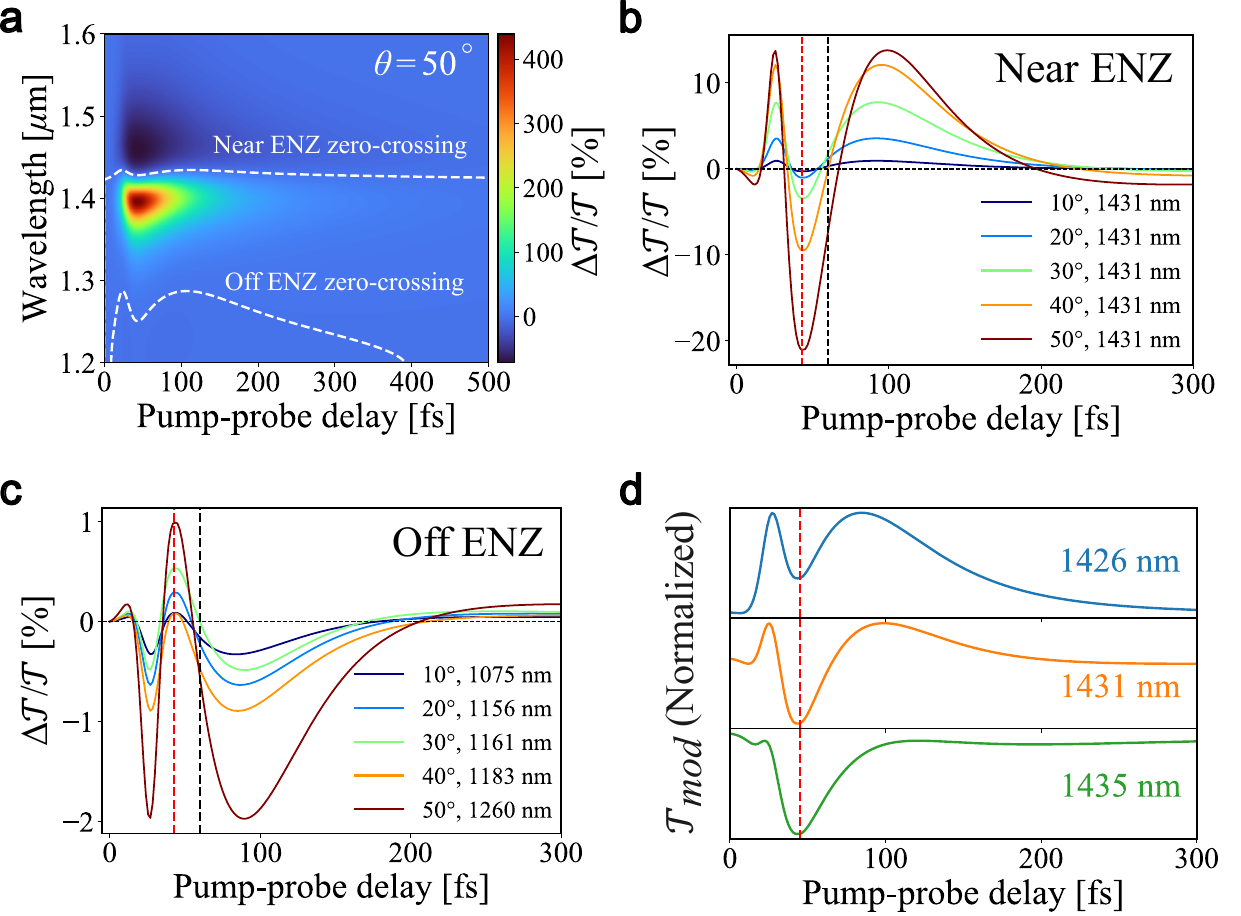}
    \caption{
    \textbf{Oscillatory transmittance dynamics of the inverse-designed cavity.}
    \textbf{a}, Relative transmittance response as a function of pump–probe delay and probe wavelength, with dotted lines marking the near ($\sim$1425\,nm) and off-ENZ ($\sim$1200--1300\,nm) zero-crossing lines. The probe incident angle is $50^\circ$. 
    \textbf{b}, Near-ENZ ($\sim$1425 nm) transmittance oscillations for various probe incident angles, exhibiting three modulation cycles with $\sim$20\,fs durations. The red dotted line marks the maximum electron temperature ($\sim$45 fs), and the black dotted line denotes 60\,fs, highlighting the duration of the modulation-cycles.
    \textbf{c}, Off-ENZ ($\sim$1200--1300\,nm) transmittance oscillations exhibiting opposite sign behavior compared to the near-ENZ case, while preserving a similar temporal timescale. 
    \textbf{d}, Temporal evolution of the modulated transmittance ($\mathcal{T}_{\mathrm{mod}}$) at probe wavelengths near the ENZ region, highlighting the distinct temporal profiles of the sub-20\,fs features. The red dotted line marks the time of maximum electron temperature ($\sim$45\,fs).
    }
    \label{fig:fig_3}
\end{figure}

Figure~\ref{fig:fig_3}(a) presents the transmittance dynamics for a $p$-polarized probe incident at $50^\circ$, with the pump incident at $60^\circ$. Two distinct spectral regions exhibit oscillatory transmittance, appearing both near the TCO ENZ wavelength and away from the ENZ region, consistent with the prediction in Figure~\ref{fig:fig_2}(b). As shown in Figure~\ref{fig:fig_3}(b), the modulation sign reverses twice before the electron temperature reaches its peak value (red dotted line), reflecting the ultrafast reversal of dominance between refraction ($\alpha \Delta \varepsilon'$) and absorption ($\beta \Delta \varepsilon''$). This competition drives oscillatory switching between relatively metallic and dielectric optical states on timescales of a few optical cycles. Notably, the near-ENZ and off-ENZ spectral regions exhibit opposite sign-reversal trends while maintaining similar temporal durations (Figure~\ref{fig:fig_3}(b,c)). This inversion originates from opposite sign combinations of the refractive ($\alpha$) and absorptive ($\beta$) sensitivities in the two regions: $\alpha < 0$ and $\beta > 0$ near ENZ, but $\alpha > 0$ and $\beta < 0$ off ENZ (Supplementary Figure~6). While this balance occurs in distinct spectral ranges near and off ENZ, the ENZ enhancement produces a substantially stronger modulation amplitude of the oscillations. In particular, the modulation amplitude is approximately one order of magnitude larger near the ENZ wavelength than in the off-ENZ region (Figure~\ref{fig:fig_3}(b,c)). The amplitude further increases with probe incidence angle, whereas the modulation-cycle duration, defined as the 0--peak--0 interval between successive zero-crossings, remains nearly unchanged. This angular dependence enables efficient amplitude control without sacrificing modulation speed. 

The contrast is further reflected in the spectral position of the oscillatory response. The near-ENZ oscillations remain spectrally fixed at 1431\,nm, whereas the off-ENZ response redshifts with increasing probe incidence angle. This behavior arises because the material-defined ENZ condition pins the near-ENZ oscillatory response, whereas the off-ENZ balance depends more strongly on the incidence angle due to the absence of ENZ enhancement. More importantly, the oscillatory dynamics, both near and away from the ENZ wavelength, reach a modulation speed of $\sim$20\,fs per cycle, corresponding to roughly 4--5 optical cycles of the 1260 and 1431\,nm probe beams. This places the response in a regime where non-adiabatic effects may arise~\cite{Moti_TCO_experiment}, thereby motivating further theoretical investigation. Notably, a two-temperature framework coupled with a full Maxwell solver has previously been used to predict related time-refraction phenomena~\cite{TTM_FDTD}. Supplementary Section~11 provides additional transmittance responses for different probe angles and polarizations, further illustrating the tunability of the oscillatory transmittance dynamics.

Figure~\ref{fig:fig_3}(d) shows the normalized transmittance response near the ENZ probe wavelengths, revealing the few optical-cycle-timescale (sub-20\,fs) temporal features whose shapes vary with probe wavelength. While similar signatures have been experimentally observed in earlier studies~\cite{Chang_Nature_photonics, initial_dynamics_1, initial_dynamics_2, Moti_TCO_experiment}, our prediction exhibits greater complexity in the initial temporal regime before the electron temperature reaches its peak value (red dotted line), with significantly shorter characteristic timescales. This additional complexity and the corresponding shortening of the temporal features arise from the extreme electron temperatures achieved in the TCO through the inverse-designed cavity.

An important consideration is that the sensitivity of the Drude damping ($\Gamma$) to electron temperature plays a central role in determining both the magnitude and spectral extent of the oscillatory response. A steeper dependence generally strengthens the oscillation and broadens its spectral range, reflecting a stronger competition between refractive and absorptive changes in the optical character of the material. This, in turn, provides intuitive guidance for microstructural control: higher impurity concentrations, greater lattice disorder, smaller grain sizes, and increased roughness may reinforce the oscillatory response and extend its operational range. A more complete understanding of how the different electron momentum-relaxation channels in TCOs can be controlled is therefore essential, motivating dedicated studies of their microscopic origins and electron temperature dependence~\cite{Nanophotonics_scattering, Scattering_linear}.

\subsection*{Optical-cycle-timescale $\Delta n$ oscillations via thermionic emission}

\begin{figure}[htbp]
  \centering
  \includegraphics[width=1\textwidth]{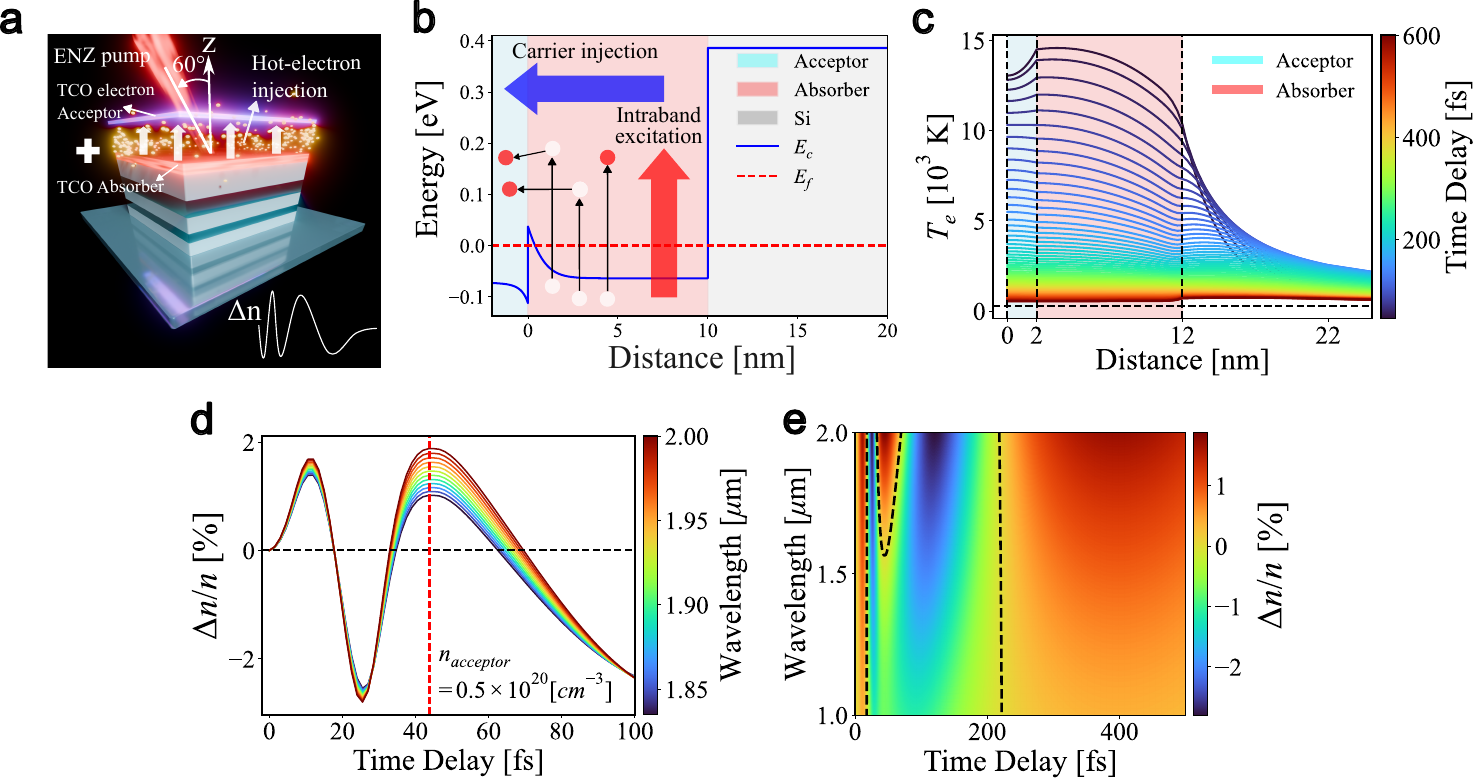}
  \caption{
  \textbf{$\Delta n$ oscillations in the TCO electron-acceptor layer under thermionic carrier injection.} 
  \textbf{a}, Illustration of the multilayer structure incorporating the bilayer TCO configuration. A 2\,nm low-doped TCO electron-acceptor layer is placed on top of the TCO absorber layer. Except for the added acceptor layer, the geometry is identical to that of the inverse-designed cavity in Figure~\ref{fig:fig_1}(b). The hot carriers generated in the absorber layer are transferred to the acceptor layer via thermionic injection. The absorber and acceptor layers are in direct physical contact; the schematic gap is included only for conceptual illustration. 
  \textbf{b}, Conduction band alignment of the bilayer TCO structure, highlighting the low barrier height and narrow barrier width that facilitates efficient carrier injection into the acceptor layer upon intraband optical pumping.
  \textbf{c}, Simulated spatiotemporal evolution of electron temperature along the vertical axis of the multilayer cavity, showing a significant temperature rise in the acceptor layer required for the oscillatory $\Delta n$ response. The line color indicates the time delay after pump excitation. 
  \textbf{d}, $\Delta n/n$ of the acceptor layer, exhibiting an oscillatory response with three modulation cycles of $\sim$20\,fs duration and a peak modulation amplitude of $\sim$2\%.
  \textbf{e}, Two-dimensional map of the $\Delta n/n$, showing the spectral range of the oscillatory response. The dotted lines mark the zero-crossing regions of the oscillatory response.
  }
  \label{fig:fig_4}
\end{figure}

Here, we aim to establish a realistic platform for achieving $\Delta n$ oscillations on few-optical-cycle timescales, with the understanding from the Sommerfeld model that oscillations are driven under high electron temperatures by substantial carrier density modulation (Figure~\ref{fig:fig_1}(d)). To theoretically realize the electron-temperature-driven carrier-density modulation required for $\Delta n$ oscillations in an experimentally accessible structure, we place a 2-nm-thick, low-doped TCO layer on top of the inverse-designed cavity (Figure~\ref{fig:fig_4}(a)). This additional layer enables interfacial hot-carrier injection via thermionic emission, allowing simultaneous modulation of carrier density and effective mass. This thickness is experimentally achievable using ALD and RF magnetron sputtering~\cite{ITO_ALD_1nm, ITO_sputtering_1nm}, and its strong spatial confinement enables ultrafast thermalization of the thermionically injected electrons. The carrier density of this acceptor layer, $0.5 \times 10^{20}\,\mathrm{cm}^{-3}$, is engineered to be 15 times lower than that of the absorber layer, providing two key advantages. First, the reduced baseline carrier concentration allows substantial relative carrier-density modulation under carrier injection. Second, the low-doped TCO remains nearly transparent at the ENZ pump wavelength (1425\,nm), thereby minimizing its influence on the static optical response of the inverse-designed cavity and preserving strong absorption in the underlying TCO absorber layer. Importantly, the thickness and carrier density of the acceptor layer are selected to reproduce a comparable level of carrier-density modulation and chemical potential reduction to that predicted by the hypothetical carrier-injection scenario described by the Sommerfeld model (Supplementary Section~13). The hypothetical carrier-injection model in the previous section, therefore, provides a useful design guideline for selecting material and geometrical parameters in a more experimentally relevant configuration.

The hot-carrier injection process from the TCO absorber into the acceptor layer (Figure~\ref{fig:fig_4}(a)) can be understood as the following sequence. Under the same excitation conditions illustrated in Figure~\ref{fig:fig_1}(b), strong intraband absorption in the TCO absorber layer populates electrons into high-energy states. Rapid electron-electron scattering then establishes a strongly broadened Fermi-Dirac distribution characterized by an extreme electron temperature. A substantial fraction of carriers in the high-energy tail subsequently injects thermionically into the adjacent acceptor layer. Because the acceptor layer has an increasing electronic density of states at higher energies due to its non-parabolic conduction band, it can effectively accommodate the injected carriers. Supplementary Figure~11(b) shows the resulting increase in injected carrier density with increasing electron temperature. The electron temperature-dependent thermionic emission rate, $dN_{\text{esc}}/dt$, is calculated using a modified Richardson-Dushman equation~\cite{Thermionic_Sim,Thermionic_Sim2,Thermionic_space_charge}:
\begin{equation}
\frac{dN_{\text{esc}}}{dt}
=
A R_1 R_2 (k_{\mathrm B} T_{\mathrm e})^2
\exp\left[
-\frac{eE_f - \mu(T_{\mathrm e}) + e\phi + \phi_{\mathrm{sc}}}{k_{\mathrm B} T_{\mathrm e}}
\right],
\end{equation}
where $N_{\text{esc}}$ is the total yield of the escaped electrons, $A = 4\pi m^\ast h^{-3}$ is the renormalized Richardson constant $[\mathrm{s^{-1}\,m^{-2}\,K^{-2}}]$, $R_1$ and $R_2$ are the semi-major and semi-minor axes of the elliptical emission region, $m^\ast$ is the electron effective mass, $h$ is Planck’s constant, $E_f$ is the Fermi level, $\mu(T_{\mathrm e})$ is the temperature-dependent chemical potential, $\phi$ is the interfacial barrier height, and $\phi_{\mathrm{sc}}$ is the space-charge-induced energy barrier. The space-charge energy barrier is approximated as
\begin{equation}
\phi_{\mathrm{sc}} \approx \frac{a N_{\text{esc}} e^2}{R_1},
\end{equation}
where $a$ is a geometrical factor, often taken as $a = 16/(3\pi\epsilon_r^2)$ for uniform emission disks, and $\epsilon_r$ is the static relative permittivity of the emission region.

Assuming a square-wave temporal profile for the electron temperature over the pulse duration $\tau$, the total emitted yield can be integrated analytically as~\cite{Thermionic_space_charge}
\begin{equation}
N_{\text{esc}}
=
\frac{k_{\mathrm B} T_{\mathrm e}}{ae^2/R_1}
\log\left[
1 + A\tau\pi R_2 ae^2 k_{\mathrm B} T_{\mathrm e}
\exp\left(
\frac{-eE_f + \mu(T_{\mathrm e}) - e\phi}{k_{\mathrm B} T_{\mathrm e}}
\right)
\right],
\label{eq:N_yield}
\end{equation}
which captures the self-consistent update of the emission rate on the evolving space-charge energy barrier and chemical potential shift. Notably, under heavy doping, the interfacial depletion region becomes sufficiently narrow that carrier transfer approaches the thermionic-field-emission limit~\cite{Thermionic_field_theory}, rendering the initial barrier effectively negligible. We therefore set the initial interfacial barrier height to $\phi = 0$. Following photoexcitation, however, the effective barrier, $eE_f - \mu(T_{\mathrm e}) + e\phi$, dynamically increases due to the temperature-dependent downward shift of the chemical potential and the electrostatic contribution from accumulated transferred charge, namely the space-charge effect.

Although the present model uses a thermionic-emission boundary condition, we note that carrier injection at realistic interfaces can depend on microscopic interfacial properties and therefore deviate from the idealized boundary condition. A full treatment of this dependence lies beyond the scope of the present work. Nevertheless, prior studies of hot-carrier injection suggest that nanoscale roughness or disorder can relax momentum-conservation constraints at the interface. This allows carrier injection to be governed more strongly by the available density of states, thereby potentially improving extraction efficiency ~\cite{Khurgin_hot_electron, Khurgin_hot_electron2, Khurgin_hot_carrier}. In this context, the non-parabolic conduction band of TCOs can facilitate efficient extraction of high-energy electrons by providing a larger density of accessible states at higher electron energy levels. Because interfacial roughness and disorder can be tailored through deposition conditions~\cite{Jongbum_photodetector}, carrier extraction in realistic geometries can be further optimized through process control.

The two-temperature model is employed to estimate the evolution of electron temperature in the acceptor layer following hot-carrier injection. Figure~\ref{fig:fig_4}(c) shows the calculated spatiotemporal evolution of the electron temperature along the out-of-plane direction of the multilayer stack. The calculation shows that the acceptor layer reaches an electron temperature comparable to that of the absorber, owing to strong confinement in the 2\,nm TCO layer. This pronounced temperature buildup in the acceptor layer, together with the substantial increase in carrier density, establishes the conditions under which the oscillatory response emerges. In this deeply confined regime, thermalization is rapidly achieved because the electron-electron scattering rate $\tau_{ee}^{-1}$ increases with both electron temperature and carrier energy~\cite{LEDD, LEDD_rigorous}. More specifically, although the conventional Gurzhi formula is no longer directly applicable to TCOs, $\tau_{ee}^{-1}$ has been shown to increase approximately linearly with electron temperature~\cite{LEDD}. We therefore calculate the $\tau_{ee}$ from the phenomenological model in Ref.~\cite{TTM_FDTD}, which qualitatively agrees with the more rigorous treatment in Ref.~\cite{LEDD}, and estimate the scattering time of the carrier distribution to be within 5--6\,fs (Supplementary Section~7). The two-temperature model therefore provides a reasonable approximation of the overall dynamics after the initial electron thermalization period, which is expected to occur within a few scattering events, corresponding to approximately 10--20\,fs. Although the non-thermal electron contribution may affect the earliest response during the first tens of femtoseconds, it is expected to play a secondary role in determining the qualitative trends of the overall dynamics. This treatment is supported by prior studies predicting relatively weak non-thermal contributions in TCOs and plasmonic metals compared with the thermalized-electron response~\cite{TTM_FDTD, LEDD, LEDD_rigorous, Gold_NP_nonthermal}. Nevertheless, developing a more rigorous treatment to capture the contribution of non-thermal electrons during the earliest stage of the oscillatory dynamics remains a fundamentally important direction for future investigation, although it lies beyond the scope of the present work. 

Importantly, while the present results are limited to the dynamics of thermalized electrons, even these thermal contributions remain incompletely understood in the extreme electron-temperature regime. The present results show that oscillatory ultrafast modulation can readily emerge from thermalized-electron dynamics alone. This observation is significant because such thermal electron-induced oscillatory behavior is often overlooked in the interpretation of TCO dynamics. The results presented here, therefore, provide a foundation for future studies to determine which aspects of experimental oscillatory responses originate from thermalized electrons and which arise from additional non-thermal or coherent nonlinear optical effects, such as multiphoton absorption and virtual transitions, that may influence the dynamics at stronger excitation levels.

We next calculate the resulting $\Delta n$ response of the electron-acceptor layer. As discussed in Supplementary Section~7, quantum-confinement features in the 2-nm-thick TCO layer are expected to be strongly broadened, validating the use of the bulk-like Drude response~\cite{khurgin_quantization}. The $\Delta n$ response of the electron-acceptor layer can therefore be calculated using the framework described by Eqs.~\ref{eq:updated_Drude_index}--\ref{eq:damping_update}. The acceptor layer exhibits an oscillatory $\Delta n$ response comprising three modulation cycles with a characteristic duration of $\sim$20\,fs, corresponding to approximately three optical cycles for a probe beam in the 1.8--2.0\,$\mathrm{\mu m}$ range. The dotted lines in Figure~\ref{fig:fig_4}(e) indicate the zero-crossing boundaries that define the oscillatory region. It is noteworthy that the electron-temperature sensitivity of the Drude damping ($\Gamma$), which is strongly microstructure-dependent, can further influence both the spectral range and amplitude of the oscillations, as discussed in the following section. Altogether, these results identify thermionic carrier injection in bilayer TCO structures as a promising platform for emerging time-varying photonic applications operating on a few optical-cycle timescale. In the following section, we demonstrate that this response can be further tailored toward sub-optical-cycle timescales.

\subsection*{Engineering $\Delta n$ oscillations toward sub-optical-cycle timescales}

Having demonstrated the oscillating $\Delta n$ responses in the TCO acceptor layer under thermionic injection and extreme electron-heating conditions, we now show that these dynamic behaviors can be readily controlled. The duration and spectral range of each modulation cycle in Figure~\ref{fig:fig_4}(d) can be further reduced by tailoring the acceptor-layer carrier density, employing shorter pump pulses, or controlling the peak electron temperature.

\begin{figure}[htbp]
  \centering
  \includegraphics[width=0.5\textwidth]{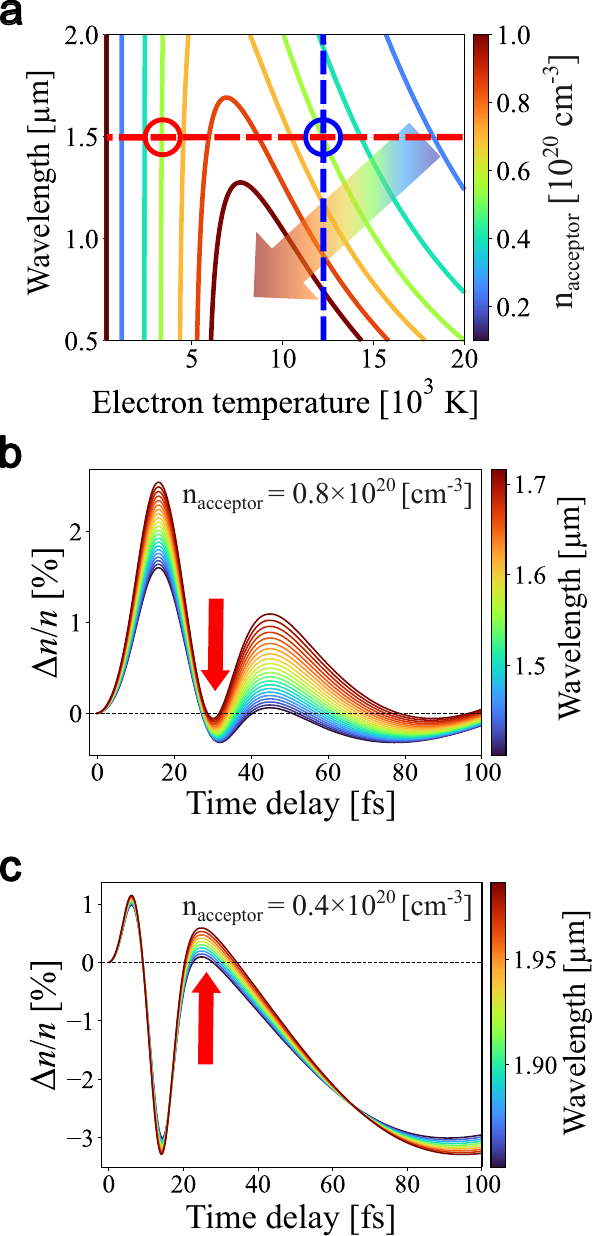}
  \caption{
  \textbf{Spectral and temporal engineering of the $\Delta n$ oscillation.} 
  \textbf{a}, Zero-crossing boundaries of $\Delta n(T_\mathrm{e})$ for a range of acceptor layer initial carrier density $n_{\mathrm{acceptor}}$, demonstrating a broad spectral tunability (See text for description of lines and circles). 
  \textbf{b}, Demonstration of $\Delta n/n$ tunability achieved by modifying the acceptor layer carrier density ($0.8 \times 10^{20}\,\mathrm{cm}^{-3}$) and peak electron temperature, resulting in four modulation cycles within 100\,fs with sub-optical-cycle timescale features (red arrow).
  \textbf{c}, $\Delta n/n$ for $n_{\mathrm{acceptor}} = 0.4 \times 10^{20}\,\mathrm{cm}^{-3}$, and shortened pulse excitation, highlighting the further reduction of the modulation timescale.
  }
  \label{fig:fig_5}
\end{figure}

We first examine this spectral and temporal tunability of the oscillations by plotting the zero-crossing boundaries of the $\Delta n$ sign map---analogous to the boundaries shown in Figure~\ref{fig:fig_2}(a)---for the acceptor layer over a range of carrier densities. Each boundary in Figure~\ref{fig:fig_5}(a) represents the wavelength and electron temperature ranges where the sign of $\Delta n$ reverses, with the color bar indicating the corresponding acceptor layer carrier density. As the carrier density increases, following the colored arrow, the contour peak shifts to shorter wavelengths and simultaneously becomes narrower. This blueshift indicates that the oscillatory regime can be pushed toward the visible by increasing the carrier density. This trend can be understood by imagining a vertical line at a fixed peak electron temperature: the wavelength at which this line first intersects the contour at a given carrier density (blue circle) gives the spectral range at which oscillatory behavior can occur. At the same time, narrowing the contour modifies the temporal response. Because the temperature trajectory takes longer to reach the first boundary crossing (red circle), the first modulation cycle becomes longer, whereas the reduced separation between successive crossings (between red and blue circles) shortens the second cycle. The corresponding time-domain response for different carrier concentrations is examined in the preceding section (Figure~\ref{fig:fig_6}(a)).

To demonstrate the potential of this oscillatory $\Delta n$ mechanism for achieving modulation on the scale of a single optical cycle, Figure~\ref{fig:fig_5}(b,c) shows representative engineered time-domain responses obtained by tuning the carrier concentration, peak electron temperature, and pump pulse duration. By setting the carrier concentration to \(0.8 \times 10^{20}\,\mathrm{cm}^{-3}\) and engineering the peak electron temperature (controllable via pump energy and incident angle), we obtain four modulation cycles within 100\,fs, over a broad spectral range of 1.4--1.7\,$\mathrm{\mu m}$ (Figure~\ref{fig:fig_5}(b)). Notably, the second modulation cycle enters the sub-optical-cycle regime (red arrow in Figure~\ref{fig:fig_5}(b)), and its duration can be tailored through the probe wavelength to values even below 1\,fs, although at the expense of modulation amplitude and a longer duration in the successive modulation cycle. Alternatively, the overall duration of the full response can be reduced by shortening the pump pulse. Figure~\ref{fig:fig_5}(c) shows the $\Delta n$ response upon 5\,fs pump excitation with 2\,mJ/cm$^2$ fluence, with a acceptor layer carrier concentration of \(0.4 \times 10^{20}\,\mathrm{cm}^{-3}\). Remarkably, the first modulation period reaches a duration of 9\,fs, followed by two additional modulation cycles within the $\sim$10\,fs range. However, the model validity under shortened pump durations requires further investigation, and the resulting modulation trend obtained through pump-duration control should be treated as qualitative rather than fully quantitative. These results highlight carrier concentration, electron temperature, and pump pulse duration as effective parameters for tailoring the spectral and temporal characteristics of the oscillatory responses, enabling flexible control of refractive index modulation on a few to sub-optical-cycle timescales.

\section*{Sensitivity analysis of the oscillatory $\Delta n$ response}

\begin{figure}[htbp]
    \centering
    \includegraphics[width=1\textwidth]{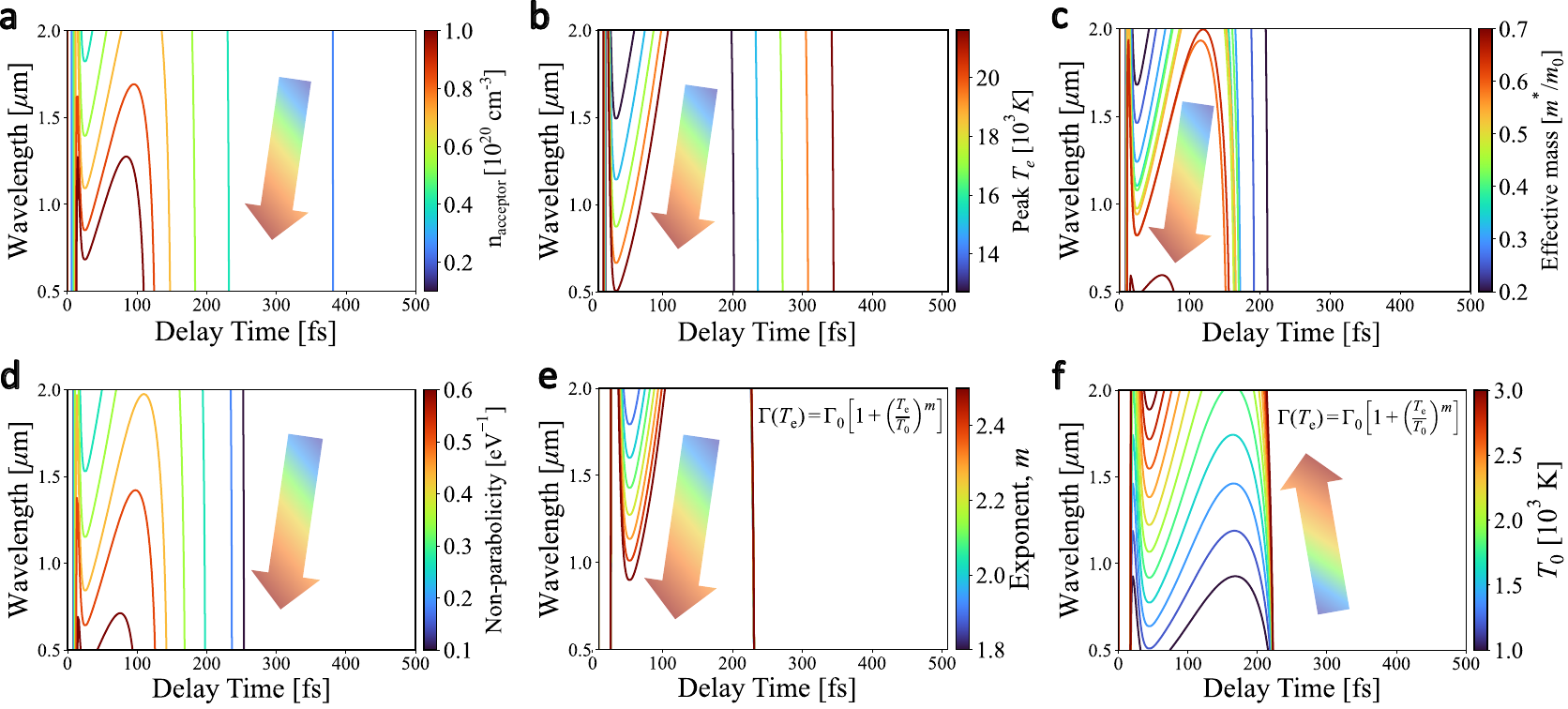}
    \caption{
    \textbf{Sensitivity analysis of the zero-crossing contours of oscillatory $\Delta n$ response in the TCO acceptor layer.}
    \textbf{a}, Zero-crossing contours of $\Delta n$ as a function of probe wavelength and time with different acceptor layer carrier densities. The colored arrow indicates the direction of the shift in the oscillatory response as the corresponding colorbar parameter varies.  
    \textbf{b}, Zero-crossing contours of $\Delta n$ for varying peak electron temperatures in the acceptor layer.  
    \textbf{c}, Zero-crossing contours of $\Delta n$ as a function of the conduction band minimum effective mass, $m^*_0$.  
    \textbf{d}, Zero-crossing contours of $\Delta n$ as a function of the non-parabolicity parameter.
    \textbf{e}, Zero-crossing contours of $\Delta n$ for different values of the Drude-damping power-law exponent $m$.  
    \textbf{f}, Zero-crossing contours of $\Delta n$ for different values of the characteristic temperature parameter $T_0$ in the Drude-damping model
    }
    \label{fig:fig_6}
\end{figure}

In this section, we present a sensitivity analysis of the oscillatory response to variations in the material parameters and the peak electron temperature. This analysis highlights the broad accessibility of the predicted oscillatory behavior and demonstrates the potential to tailor the dynamic response through additional material design and excitation degrees of freedom. For instance, the zero-crossing regions can be systematically engineered by changing the carrier concentration of the acceptor layer, as discussed in the previous section. In Figure~\ref{fig:fig_6}(a), increasing the carrier concentration blueshifts the zero-crossing regions toward the visible range and shortens the duration, in agreement with the trend predicted in Figure~\ref{fig:fig_5}(a). Figure~\ref{fig:fig_6}(b) further shows that the peak electron temperature provides an additional tuning parameter at a fixed acceptor-layer carrier concentration of $n_{\mathrm{acceptor}} = 0.5 \times 10^{20}~\mathrm{cm}^{-3}$. Increasing the peak electron temperature extends the zero-crossing region toward shorter wavelengths, spectrally extending the occurrence of the oscillatory $\Delta n$ response from the near-infrared into the visible. Beyond demonstrating the controllability, the electron temperature sweep accounts for possible uncertainties in the absolute electron temperature reached under experimental conditions. Although the exact electron temperature varies with the band structure, density of states, electron heat capacity, and electron–phonon coupling strength of individual samples, the sweep analysis shows that the qualitative oscillatory response remains robust over a broad electron-temperature range. In addition, the boundary positions can be further refined through experimental fitting. For instance, the experimental zero-crossing boundaries can be extracted from pump-intensity-dependent measurements, providing an experimentally informed extension of the present theoretical framework.

The spectral position of the zero crossing can also be tailored through band structure engineering, for example, achieved by controlling microstructure and defects during material growth and processing. Such modifications affect the non-parabolic conduction band, characterized by the conduction-band-edge effective mass $m^*$ and the non-parabolicity parameter $C$. Their influence on the zero-crossing is shown in Figure~\ref{fig:fig_6}(c,d). As the effective mass increases from $0.2\,m_0$ to $0.6\,m_0$, the zero-crossing shifts toward shorter wavelengths and becomes narrower in time. At $0.7\,m_0$, the response shifts abruptly into the visible range. Figure~\ref{fig:fig_6}(d) shows that increasing the non-parabolicity parameter $C$ at a fixed conduction-band-edge effective mass, $m^* = 0.24\,m_0$, likewise drives a blueshift of the zero-crossing region. This behavior originates from the stronger increase in effective mass at higher electron energies.

The same sensitivity extends to the scattering model. The oscillatory behavior depends strongly on how the Drude damping evolves with electron temperature, parameterized here by the power-law exponent $m$ and the characteristic temperature scale $T_0$ in Eq.~\ref{eq:damping_update}. The wide range of $m$ values considered reflects the substantial microstructural variations commonly found in TCO films. As shown in Figure~\ref{fig:fig_6}(e,f), larger $m$ shifts the response toward shorter wavelengths, whereas a larger $T_0$ produces the opposite trend. Taken together, these results show that the oscillatory response persists across a broad range of material parameters, band-structure variations, scattering behaviors, and excitation conditions. The physical picture developed here is therefore not restricted to a single material parameter set, but provides a more general framework for understanding and engineering optical-cycle-timescale dynamics in TCO systems. As a result, this sensitivity analysis points toward a design strategy in which material processing, band-structure engineering, and excitation conditions are jointly optimized to control non-trivial optical responses on timescales of a few optical cycles.

\section{Discussion \& Outlook}\label{sec4}
By demonstrating a few optical-cycle-timescale modulations in TCOs, this work highlights the connection between fundamental studies of thermalized-carrier dynamics and the device-level implementation of emerging time-varying photonic media. Here, we briefly outline promising future studies that build on the insights of this work, spanning fundamental studies of non-thermal carrier dynamics, the emerging physical concept of intervalley transfer dynamics, and metasurface-based implementations of photonic time crystals using the carrier injection framework. 

While ultrafast scattering dynamics justify the rapid thermalization in TCO layers, a comprehensive theoretical treatment of the dynamics involving non-thermal electrons remains an important direction for future research. In particular, Ref.~\cite{LEDD_rigorous} presents a rigorous framework for modeling TCO dynamics using the Boltzmann equation beyond the relaxation-time approximation. In parallel, self-consistent electrodynamic approaches that couple Maxwell’s equations to a temporally varying permittivity have been developed to capture field propagation in time-varying media~\cite{LEDD_rigorous, TTM_FDTD}. Combining non-thermal carrier evolution, field propagation, and carrier injection-driven modulation within a unified framework would enable a more complete and accurate description of the overall dynamics.

While the present work focuses on spatial carrier transfer across different TCO layers, a potentially richer dynamic response could arise from the intervalley carrier transfer in the momentum space. Under strong ENZ-enhanced pumping conditions relevant to TCOs, the electron temperature can exceed the Fermi temperature by a large margin, populating states far above the band edge and enabling intervalley transitions across multiple conduction-band valleys. Such a transfer can lead to an abrupt, non-monotonic change in effective mass, and therefore result in a rapid, unconventional modification of the plasma frequency. Capturing this behavior requires a more rigorous treatment incorporating the full band structure, intervalley scattering terms, and valley-dependent dipole matrix elements, which is well beyond the scope of the present work. Nevertheless, exploring this momentum-space carrier-transfer dynamics is a fundamental and unexplored aspect of TCO nonlinearities and therefore offers a compelling direction for future work.

On the device level, resonant metasurfaces have been shown to relax the stringent refractive index modulation required for photonic time crystals. Specifically, theoretical demonstrations using Si nanosphere arrays have shown that photonic time crystal behavior can emerge from as little as a $1\%$ $\Delta n/n$, once an optical-cycle-timescale refractive index modulation is achievable~\cite{Expanding_momentum_nature, PTC_science}. However, $1\%$ modulation already approaches the upper limit in conventional nonlinear materials such as Si, where practical relaxation times are several orders of magnitude slower than the optical-cycle regime~\cite{Si_threshold, Si_threshold2, theory_application_PTC}. To this end, our TCO thermionic carrier injection scheme, which achieves $\sim2\%$ refractive index modulation at a few optical-cycle speeds, may represent a significant step toward realizing the building blocks of resonant metasurface-based photonic time crystals.

Taken together, our numerical investigation based on the two-temperature model suggests a plausible explanation for the unusual oscillatory optical response observed in TCOs by capturing the non-trivial dynamics that were previously unexplained. While we do not claim that this is the only possible explanation---intervalley transfer dynamics and interband-related nonlinear effects at even higher pump intensities deserve further investigation---we aim to draw attention to this interesting problem and stimulate further studies toward a more definitive understanding and, ultimately, practical applications. Through electron temperature-dependent analysis on dynamic optical response, we demonstrate that the transient response can be broken down into multiple cycles of sign reversal under extreme electron temperatures, enabling dramatically enhanced modulation speeds. For this purpose, an inverse-designed cavity was optimized to maximize electron temperature in an ultrathin TCO layer. The calculated transmittance response exhibited oscillatory dynamics with approximately $20\%$ modulation at probe wavelengths of 1.26 and 1.431\,$\mathrm{\mu m}$, with a period of $\sim$20\,fs, corresponding to approximately 4--5 optical cycles of the probe beam. The observed oscillations in transmittance were shown to originate primarily from a time-dependent shift in the balance between enhanced absorption and modified refraction, resulting from the competition between reduced plasma frequency and increased Drude damping. Importantly, the oscillatory response was found to be more closely related to modulation of the extinction coefficient ($\Delta k$) than to that of the refractive index ($\Delta n$).

We further examined whether optical-cycle-timescale modulation can be realized in the refractive index, which is a more fundamental optical quantity. Extreme electron temperatures were found to be essential for enabling the $\Delta n$ oscillations. Analysis based on hypothetical carrier injection showed that dynamic carrier-density modulation in TCOs drives $\Delta n$ oscillations through competition between $n_e$ and $m^*$. To realize this mechanism in a practical platform, we numerically investigate a thermionic carrier-injection scheme in which a 2-nm-thick TCO electron-acceptor layer is added on top of the inverse-designed cavity. The acceptor layer exhibited an oscillatory $\Delta n$ of approximately $2\%$ at 1.8--2.0\,$\mathrm{\mu m}$ with a characteristic duration of $\sim$20\,fs, corresponding to approximately three optical cycles of the probe beam. Furthermore, we examined the sensitivity and broad tunability of the $\Delta n$ oscillations in modulation speed, spectral range, and amplitude by tailoring material and excitation parameters, including carrier concentration, band structure, scattering dynamics, and pump conditions. Through this parameter space, we further demonstrated $\Delta n$ on timescales shorter than a single optical cycle. Altogether, this work not only suggests a plausible mechanism underlying the non-trivial femtosecond response in TCOs but also establishes a practical route toward realizing optical-cycle-timescale refractive index modulation through thermionic carrier injection in TCOs, with important implications for emerging photonic time crystals.

\section*{Methods}
\subsection*{Inverse design and static optical response}
The optical response of the multilayer structures was calculated using the transfer-matrix method. The inverse-designed cavity was optimized by gradient-based backpropagation combined with the transfer-matrix method to maximize absorption in the ultrathin TCO layer at the target ENZ wavelength. 

\subsection*{Transient optical response}
The resulting spatially resolved absorbed-energy profile from the transfer-matrix method was then used as the heat source in a two-temperature model to calculate the transient electron temperature dynamics following ultrafast optical excitation. The transient optical properties of the TCO film were obtained using an electron temperature-dependent Drude model incorporating variations in plasma frequency, chemical potential, and Drude damping. For the TCO bilayer structure, hot-carrier transfer across the interface was modeled using a thermionic-emission boundary condition based on a modified Richardson–Dushman formulation. The resulting electron temperature and changes in carrier density were used to calculate the transient refractive index and transmission dynamics. Full model formulations, electron temperature-dependent material parameters, and comparison with experimental results are provided in the Supplementary Information.

\section*{Data Availability}
The data generated in this study are provided in the article and Supplementary Information and can be reproduced using the Python code published on Figshare under DOI: https://doi.org/10.6084/m9.figshare.33283908.

\section*{Code Availability}
The code used to generate and analyze the numerical results presented in this study has been deposited in the Figshare repository under DOI https://doi.org/10.6084/m9.figshare.33283908.

\bibliographystyle{naturemag}
\bibliography{Reference_Main}

\section*{Acknowledgement}
We thank Mordechai Segev and Ohad Segal of the Technion--Israel Institute of Technology for helpful discussions and for sharing experimental transmission-dynamics data for ITO.

\section*{Funding Statement}
The authors disclose support for the research of this work from the U.S. Department of Energy, Office of Science, Office of Basic Energy Sciences, Division of Materials Sciences and Engineering under Award Number DE-SC0017717 (transparent conducting oxides characterization and optical properties), and DEVCOM Army Research Laboratory through the collaborative project “Ultrafast Space-Time Photonics and Single-Photon Optical Modulators.”

\section*{Author Contributions}
J.I.C. conceived the project concept, developed the theoretical framework, performed numerical simulations, analyzed the data, and wrote the manuscript.
V.M. advised on the development of the theoretical models and contributed to the interpretation of the results. 
C.F. advised on conceptual development and contributed to manuscript writing. 
J.B.K. contributed to the theoretical development and to shaping the broader outlook of the work.
A.V.K., V.M.S., and A.B. supervised the research direction. 
All authors discussed the results and contributed to the final manuscript.

\section*{Competing Interests}
The authors declare no competing interests.


\end{document}